# Rare-Region Onset of Superconductivity in Granular Systems

Malcolm Durkin[1], Sarang Gopalakrishnan[3], Rita Garrido-Menacho[1], Ji-Hwan Kwon[2], Jian-Min Zuo[2], and Nadya Mason[1]

**Affiliations:**

[1]Department of Physics and Materials Research Laboratory, University of Illinois at Urbana-Champaign, Urbana IL 61822

[2] Department of Materials Science and Materials Research Laboratory, University of Illinois at Urbana-Champaign, Urbana IL 61822

[3] Department of Physics and Walter Burke Institute, California Institute of Technology, Pasadena CA 91125

**The critical behavior of disordered systems—from metals (*1*) to magnets (*2*) and superconductors (*3*)—is often dominated by the behavior of "rare regions" of a correlated phase, which control the inception and dynamics of the phase transition. Yet, despite significant theoretical (*3,4,5*) and experimental (*6,7,8,9*) interest, there has been little direct evidence of the presence of these regions, or of their role in initiating transitions. Here, we provide direct evidence for rare-region effects at the onset of superconductivity in granular superconducting islands. By considering the strong diameter-dependence of the transition, as well as observations of large fluctuations in the transition temperature as island diameters decrease, we are able to show that superconducting order first appears in unusually large grains—i.e., rare regions—within each island and, due to proximity coupling, spreads to other grains. This work thus provides a quantitative, local understanding of the onset of correlated order in strongly disordered systems.**

Strong disorder can destroy superconductivity in films via a continuous quantum phase transition; in this case, most canonical theories predict a transition to an insulating state as the normal state resistance approaches the quantum of resistance ($R_Q \sim 6.4$ KΩ) (*10*). However, multiple experiments have instead demonstrated suppressed superconductivity at much lower film resistances and phase transitions to low-resistance metals (*11,12,13*). The main theoretical paradigm for understanding such superconductor-metal transitions (*3,4,5*) assumes that the films possess emergent inhomogeneity, i.e., they break up into locally superconducting islands in a metallic matrix. In this picture, the global onset of superconductivity is controlled by rare, anomalously superconducting regions. Although many recent experimental results have been suggestive of highly inhomogeneous superconductivity in various systems (*6,7,8,9,14*), these works did not determine the nature of the onset of superconductivity, nor directly connect this to the inhomogeneity.

Here, we provide direct evidence for rare-region effects at the onset of superconductivity in granular systems having $R \ll R_Q$. We demonstrate that evaporated niobium (Nb) is a realization of a proximity coupled small grain system, and show that its critical behavior can be probed by measuring transport as a function of diameter of a mesoscopic island. We find that $T_c$ is strongly suppressed at island diameters of over 1 μm. This behavior is unexpected given conventional theories of superconductivity in small grains (*15*), as well as experiments on isolated superconducting grains (*16,17,18*), which predict suppression at 100 times smaller length scales. However, the anomalous size-dependence we observe can be explained quantitatively as a rare region effect: specifically, by a model in which the onset of superconductivity on an island coincides with the transition temperature of its largest constituent grain. This "extremal-grain" model accounts for both the size-dependence of the transition temperature and its sample-to-sample fluctuations. The present work thus provides a *local* understanding of the onset of superconductivity in highly inhomogeneous systems near a superconductor-metal transition.

Islands are composed of 70 nm thick electron-beam evaporated Nb, on top of insulating $SiO_2$ substrates, and have diameters varying between 200 nm and 2600 nm. Nb is chosen because it forms nanoscale grains when either sputtered or evaporated, with structure and grain size dependent on deposition parameters (*19*). Figure 1 shows typical TEM images of our Nb, where black crystals (the "grains") are surrounded by gray, amorphous-like material. TEM, XRD, and AFM studies confirm that the samples consist of highly conducting Nb grains embedded in less conductive Nb (Figure 1 and Supplement B). Previous studies are also consistent with evaporated Nb forming crystalline superconducting grains ($T_c \sim 9K$) surrounded by amorphous Nb having strongly suppressed superconductivity *(13,20)*. The Nb films have an exponential distribution of grain diameters, a mean grain radius of 2.3 nm and a small number of large grains having radii over 20 nm (Figure 3a); we find that the grain size distribution does not depend upon island diameter (see Supplement A). The approximate grain density extracted from the fit in Figure 3a is 43 grains per $100nm^2$. Islands are contacted in a 4-point configuration using 10nm/1nm Au/Ti leads [inset, Fig. 2(a)] and are measured below $T_c$ in a cryostat using standard lockin amplifier techniques. From film resistivity measurements, we determine a Nb Ginzburg-Landau dirty-limit coherence length of $\xi_{SC}$ ~ 27 nm (*21*).

Fig 2(a) shows typical resistance versus temperature curves for islands of various diameters, demonstrating that the islands undergo superconducting transitions to zero-resistance states and that the transition temperature depends strongly on island diameter. Fig 2(b) plots $T_c$ versus diameter for various sets of samples. Each curve represents a different Nb evaporation. The curves all show the same trend: $T_c$ is strongly suppressed as diameter is decreased below ~ 2 μm. For large islands (> 2.5 μm), $T_c$ approaches the bulk value for Nb, 9 K. However, as the island diameter is decreased, $T_c$ drops sharply, to below 4 K at 1 μm, and well below that for smaller diameters. We have measured scores of islands, and they all show the same trends; however, the curves can be shifted laterally depending on evaporation pressure (see green curve in Fig. 2b, shifted left due to a lower pressure evaporation). The black curve in Fig. 2b shows data for islands placed on underlying squares of Au (10 nm thick); the fact that this data is nearly identical to islands placed on insulators demonstrates that the normal metal (underlying, or in leads) is not suppressing the transition. The micron-length scales at which superconductivity is suppressed are far longer than other length scales related to superconductivity in Nb, such as the coherence length (~ 27 nm) or the scale at which the gap equals the discrete energy level spacing (~ 4 nm).

A phenomenological model involving extremal grains can fully account for the dependence of $T_c$ on island diameter. In this extremal grain model, the island's onset of superconductivity at $T_c$ coincides with the superconducting transition of its largest constituent grain. Larger islands have higher $T_c$ than smaller islands because they have more grains and, therefore, a higher probability of having an anomalously large, high-$T_c$ grain. The quantitative model requires two inputs. The first is the probability distribution of grain sizes, denoted $P(L)$ where $L$ is the diameter of a particular grain; this distribution is determined experimentally as $P(L)=\beta\,e^{(-\beta L)}$ with $\beta$=0.424nm$^{-1}$ (see Fig. 3a). The second input is the transition temperature, $T_c(L)$, which is given by (*2*)

$$T_c \sim T_c^0 \sqrt{L - \xi_{sc}} \qquad (1)$$

where $T_c^0$ is the bulk transition temperature and $\xi_{SC}$ is the superconducting coherence length of Nb. Since this is the transition temperature of a grain embedded in a metallic matrix, formed by amorphous Nb, superconductivity occurs when the pairing energy scale $\Delta$ is greater than the Thouless energy $E_{Th} \sim \hbar D/L^2$, where $D$ is the diffusion constant (*22*). In other words, the time an electron dwells on a grain before diffusing out, $t_{Th} = \hbar/E_{Th}$, must be longer than the time is takes to form superconducting correlations, $t_\Delta = \hbar/\Delta$. Taking the standard dirty-limit $\Delta \approx \hbar D/\xi_{SC}^2$ this criterion implies that $T_c$ is suppressed when $L \sim \xi_{SC}$ [as in Eq. (1)]. This mechanism is different from those found in superconducting grains embedded in insulators, where electrons do not diffuse out of the grain and $T_c$ is only suppressed when $\Delta$ is on the order of the single-particle level spacing of the grain.

The $T_c$'s of the islands can be compared to numerical simulation by modeling the islands as a set of $N$ grains having radii randomly generated based on P($L$). A typical island of diameter $d$ has $N \sim \rho\pi d^2/4$ grains of varying sizes, where $\rho$ is the experimentally determined grain density. The simulated distributions are consistent with those experimentally extracted from TEM images using an object finding program. The largest grain radius, $L_{max}$, can be extracted from the simulation, as shown in Fig 3(b). It is clear from this figure that the probability of an island having a grain larger than the coherence length (~ 30 nm) drops dramatically below ~ 1 μm. The island $T_c$ can then be obtained from Eq. (1) using $L=L_{max}$ and $\xi_{SC} = 23$ nm. The result of this simulation is shown in Fig. 3(c) and fits very well to evaporations E1 and E2, which were performed using similar parameters to the TEM samples. Additionally, evaporations performed using different source conditions (E3-E6) can be horizontally scaled onto the simulation, indicating a similar trend. This provides excellent correspondence with experiment and requires no free parameters in the length scales, as they were experimentally obtained.

The extremal-grain model predicts not only the size-dependence of the typical transition temperature, but also a variance in $T_c$ among islands of the same diameter. While the probability distribution of $L_{max}$ for islands of fixed diameter is predicted to follow a Gumbel distribution (*23, 24*), which is not sensitive to $d$, fluctuations of $T_c$ are sensitive to $d$, due to the varying slope of Eq. (1). Large fluctuations in $T_c$ occur when the mean value of $L_{max}$ is on the order of $\xi_{SC}$, while $L_{max} > \xi_{SC}$ leads to minimal fluctuations in $T_c$, as most islands have at least one grain that goes superconducting near the bulk transition temperature. We experimentally observe both large $T_c$ fluctuations for islands of the same diameter, and a trend of increasing fluctuations with decreased diameter, as shown in Fig. 3(d). The simulation of expected fluctuations is shown on the same plot, and show similar trends for simulation and theory.

While the extremal grain model agrees with experiment, several alternative explanations should be considered, particularly because the island normal resistance, $R_N$, also scales with $T_c$ (Fig. 2b). One candidate explanation is that each individual grain experiences its neighbors as a resistive shunt whose magnitude decreases as the island size increases, promoting superconductivity. Theories of this kind predict a critical normal-state resistance of $R_Q$, which is orders of magnitude larger than our island resistance, and also cannot explain the $R_N$ dependence of $T_c$ in our islands. A second possibility is that the suppression of superconductivity is due to normal metal suppression from the contacts. However, single Nb islands with an underlying Au film have similar $T_c$ to islands contacted in four point configuration, making suppression from the underlying Au unlikely (this similarly obviates the possible relevance of a resistive shunt). Another concern is that the structure of the islands might depend on diameter, changing both $T_c$ and $R_N$. This is inconsistent with both TEM observations, which demonstrate that grain size distribution does not vary significantly with island diameter, and with measurements performed on small diameter island arrays coupled with underlying Au (*21*), which suggest that $T_c$ is determined by the total volume of coupled Nb. The dependence of island resistance on diameter can be best explained by transport through a highly granular material, where most of the current passes through the most conducting paths. Since fewer of these less resistive paths are available for small diameter islands, both the mean value and the variation in $R_N$ is greater for smaller diameter islands. Quantitively, the $R$ vs $T_c$ trend can be well-modeled with a resistor-network model that is consistent with the rare region effect [see Supplement C].

Our results suggest a physical picture of the local nature of the superconducting state near the superconductor-metal transition: this state is inhomogeneous and is dominated by rare regions, as suggested in Refs. (*3, 25*). By exploring mesoscopic systems, we have directly quantified the influence of rare regions on superconducting transport. We have found, remarkably, that even when grains are coupled strongly enough that the normal-state resistance is small, the superconducting transition can be captured via a model of effectively decoupled "grains." In this sense, our mesoscopic superconducting islands behave like many other strongly random quantum systems, such as high-temperature superconductors (*26,27*).

**Acknowledgements:** This work was supported by the DOE Basic Energy Sciences under DE-SC0012649. SG acknowledges funding from the Walter Burke Institute. JK and JMZ are supported as part of the Center for Emergent Superconductivity, an Energy Frontier Research Center funded by the US Department of Energy, Office of Science, Office of Basic Energy Sciences, under award number DE-AC0298CH10886. The research was carried out in part in the Frederick Seitz Materials Research Laboratory Central Research Facilities, University of Illinois. We thank Vadim Oganesyan for helpful discussions.

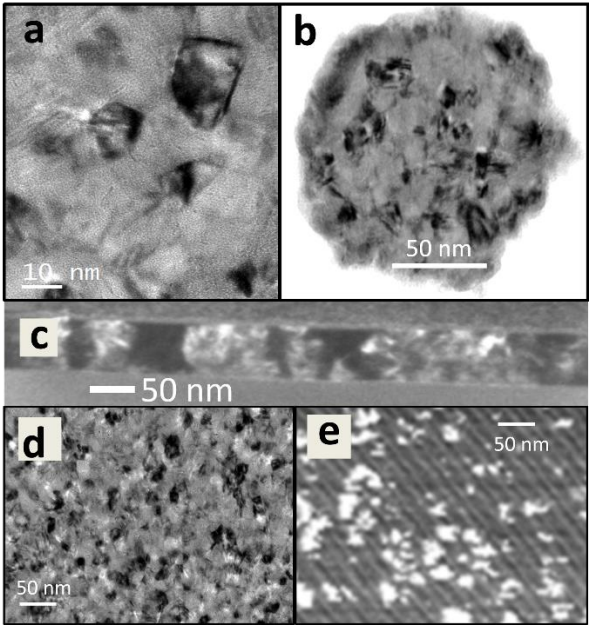

Figure 1. **TEM and Conductive AFM.** (a) Zoomed in TEM image showing crystalline Nb grains in black and amorphous-like Nb in grey. (b) TEM of a 130 nm diameter Nb island. TEM images in (a) and (b) were performed on 30 nm thick Nb. (c) Side view TEM (dark field) performed on 7 0nm thick Nb showing columnar grains. (d) TEM performed on a 2.5 μm diameter Nb island, which can be compared to (e) conductive AFM performed on 70nm thick Nb sheet. Highly conductive grains are in white (20 pA) and are separated by less conductive material shown in dark (15 pA).

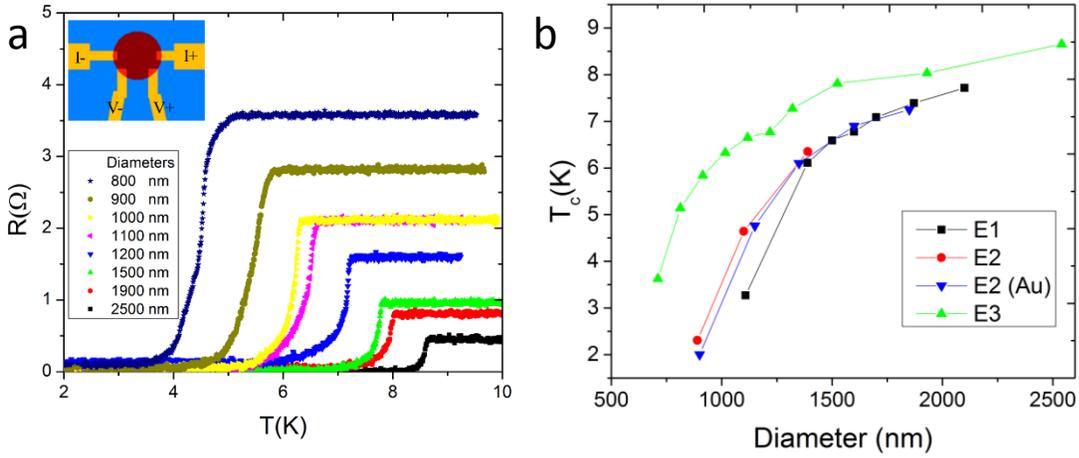

Figure 2: **Superconducting transition for different island sizes** (a) Resistance vs Temperature for various island diameters. Inset: False color optical image of island (orange) and leads (yellow). (b) $T_c$ vs Diameter for different sets of island samples. E1, E2, and E3 denote different Nb evaporations. The samples indicated by the blue triangles have underlying Au. The green curve was obtained using a lower pressure evaporation.

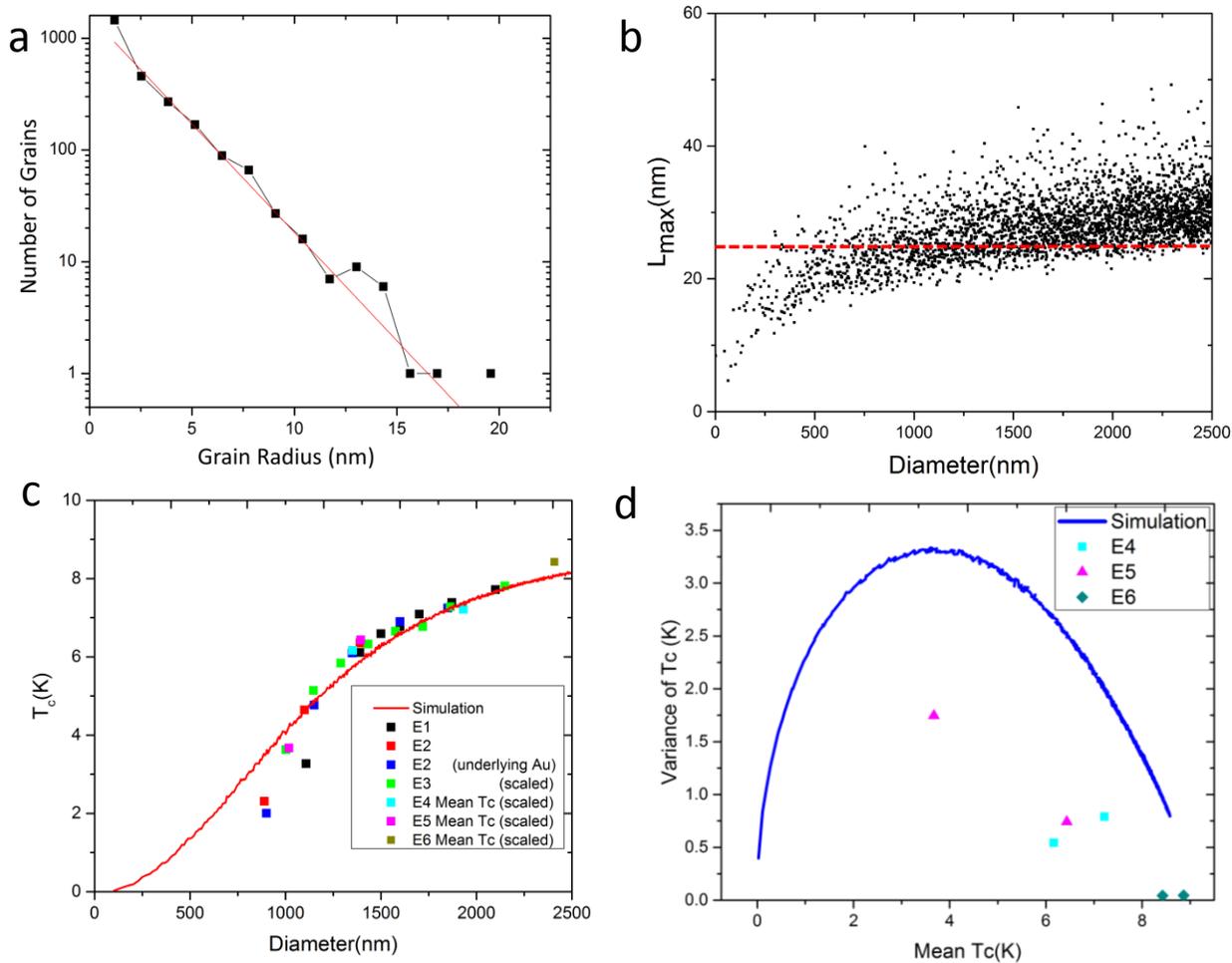

Figure 3. **Extremal grain model** (a) Grain distribution extracted from Fig.1 (d) using an object finding program. The fit shows an exponential distribution. (b) Simulated maximum grain sizes as a function of island diameter using grain distribution and density extracted from (a). The red dashed line corresponds to $\xi_{SC}$ used in (c). (c) The simulated grain sizes are applied to Eq. 1 to obtain an estimate of $T_c$. Mean simulated $T_c$ (red curve) is shown alongside data from evaporations 1 and 2 (E1 and E2), which it matches closely. Evaporations with different purities can be scaled horizontally onto the curve (E3-E6). (d) Simulated $T_c$ variance vs mean $T_c$ plotted against measurements. Each data point represents a value extracted from 5-10 islands of the same diameter.

**Supplementary Materials:**

## Materials and methods

### Materials and Methods

We pattern our samples using electron beam lithography on PMMA, electron beam evaporation, and liftoff processes on a silicon wafer with a 300nm oxide layer. We first make the four point patterns, which consist of 1nm Ti and 10nm Au, and then pattern an Nb island on top of the normal metal contacts. Nb is deposited in an ultra high vacuum (UHV) system with a 67 cm throw distance. A brief ion mill is performed in the system prior to electron beam evaporating 70 nm of Nb at a pressure of $1.0 \times 10^{-9}$ Torr. The sample is then placed in a chip carrier, contacted with a wedge bonder, and measured in a 1K cryostat using an SR830 to perform four point measurements.

### TEM Measurements of Grains

Top view TEM measurements are performed on Nb islands placed on TEM windows, which consist of a 20nm thick SiO2 membrane. The horizontal measurements are performed by cutting an island cross section using FIB and performing a side view TEM. The grain size statistics are extracted from top view TEM images using an object finder, which applies a low pass filter for smoothing and then identifies grains as areas where the image intensity is below a threshold. The threshold identification results in a large number of misidentified grains of $R < 2$nm, but correctly identifies high contrast grains larger than this. Since Fig 3(a) has high contrast regions visible in both light and dark, the mean intensity is subtracted from the image and the absolute value times negative one is taken before the threshold identification is performed, with the image processing steps shown in Fig S1 (a), (b), and (c). The result has the same grain size distribution as the dark grains, but double density of grains measured.

This object finding technique is applied on a hollow cone TEM image of Nb films, 500nm diameter Nb islands, and 200nm diameter Nb islands. While the 200nm islands did not provide a large enough sampling for an analysis, the grain distribution could be extracted from both the bulk and the 500nm islands as shown in Fig S2. The grain distribution and density of a 500nm diameter island is not significantly different from the bulk sample, indicating similar structures.

### Conductive Atomic Force Microscopy

Conductive AFM involves performing contact mode AFM measurements using a conducting cantilever, dragging the cantilever across a sample as depicted in Fig S3 (a) inset. The tip is held at constant bias, the sample is grounded at the other end of the chip, and the current is measured. Due to the narrow tip, the current measured is sensitive to the conductance near the tip and a resistor network simulation shows current peaks when the tip is above highly conductive regions [Fig SB1 (a)].

Conductive AFM measurements performed on 70nm thick Nb films find current peaks [Fig S3 (b)]. The size of these current peaks is extracted using an object finder [Fig S3 (c)] and has a comparable distribution to the grains found using TEM, suggesting that the film consists of highly conductive grains embedded in a more resistive material. To check for artifacts, the height data is taken simultaneously with the current measurements. A 3D plot with height in the Z axis shows I peaks are often accompanied by Z features such as peaks [Fig S3 (d)]. However, not all Z peaks are accompanied by I peaks and common artifacts (e.g. an increase in I when tip height is changing) are absent, suggesting that this is due to Z peaks forming at the location of grains rather than an artifact of tip interactions with the surface.

**Random Resistor Network Simulations**

The diameter dependence of the resistance can be largely explained by transport dominated by percolative paths in an inhomogeneous film. Percolative behavior has been previously of interest in the study of doped semiconductors([1],[2]) and has been studied using random square resistor networks([3]). Tuning the probability of a connection between adjacent nodes existing, $p$, and a connection not existing, $1-p$, the random network studies observe a phase transition at a critical probability, $p_c$, from finite sized clusters exist for $p<p_c$ to an infinite cluster of linked nodes throughout the network for $p > p_c$. The relevant length scale involved is the correlation length, $\xi \propto |p-p_c|^{-\alpha}$, where $\alpha$ is a scaling constant. This corresponds to the radius of the largest percolative clusters for $p < p_c$ and the radius of the largest holes in the infinite percolative cluster in $p > p_c$.

Our system corresponds more closely to the case of a good conductor in a poor conductor, which can be studied by giving the open links a large but finite resistance([4]). The resistance of this can be seen in Fig. S4 (a), with a crossover of $p_c \sim 0.6$. Sample resistance is dominated by weak links between network clusters for $p < p_c$ and conductance through a single network cluster for $p > p_c$. The finite size behavior of this model results in either a network cluster spanning the array or a hole in a network cluster dividing the array. This leads to array resistance distribution splitting as the array width, L, decreases. This splitting is shown on a logarithmic scale in Fig S4(b) at $p=0.55$. Due to proximity with $p_c$, the upper and lower curves are approximately equal in magnitude, but the lower curve is suppressed for $p<p_c$ and the upper curve is suppressed for $p>p_c$. Since we observe increasing resistance with decreasing island diameters, our data corresponds to the $p<p_c$ case involving weakly linked networks clusters.

The relevant length scale of this system is the spacing of key current paths, which corresponds to the size of a low resistance network cluster, $\xi$. These clusters do not necessarily correspond to a single grain, which would yield finite size effects on the scale of tens of nanometers. Instead, the networks could correspond to clusters of Islands, as inhomogeneities in grain density exist on the scale of hundreds of on nanometers and may explain the large increase in resistance below 700nm.

**Figures S1-S4**

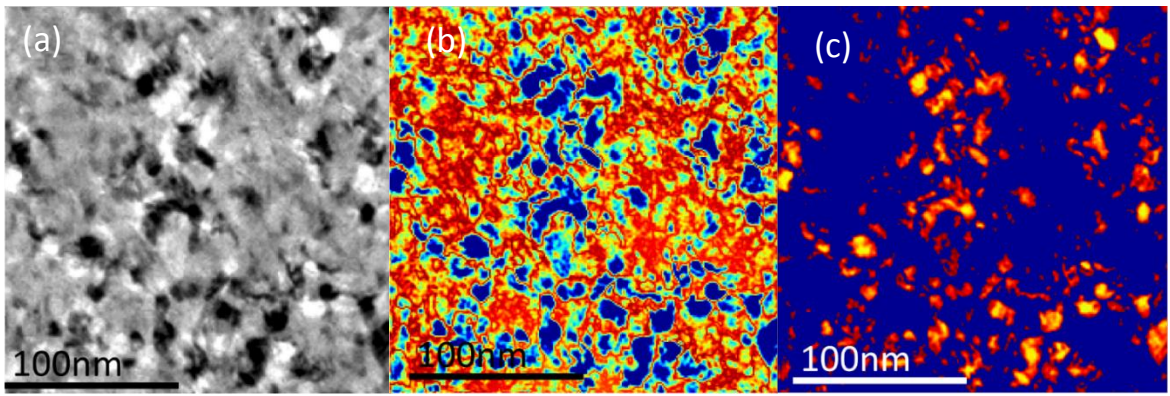

**Figure S1: TEM image processing.** (a) A raw TEM image of an Nb film. High contrast grains are visible in both black and white. (b) The TEM image after a low pass filter has been applied, mean intensity is subtracted, and the absolute value is taken. High contrast grains appear in blue. (c) The high contrast grains identified by the object finder are shown in red and orange.

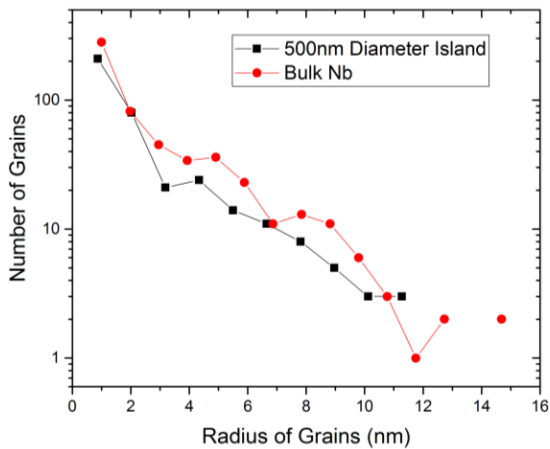

**Figure S2: TEM grain distribution.** The grain distribution of a 500nm diameter Nb island compared to a bulk sample of the same area.

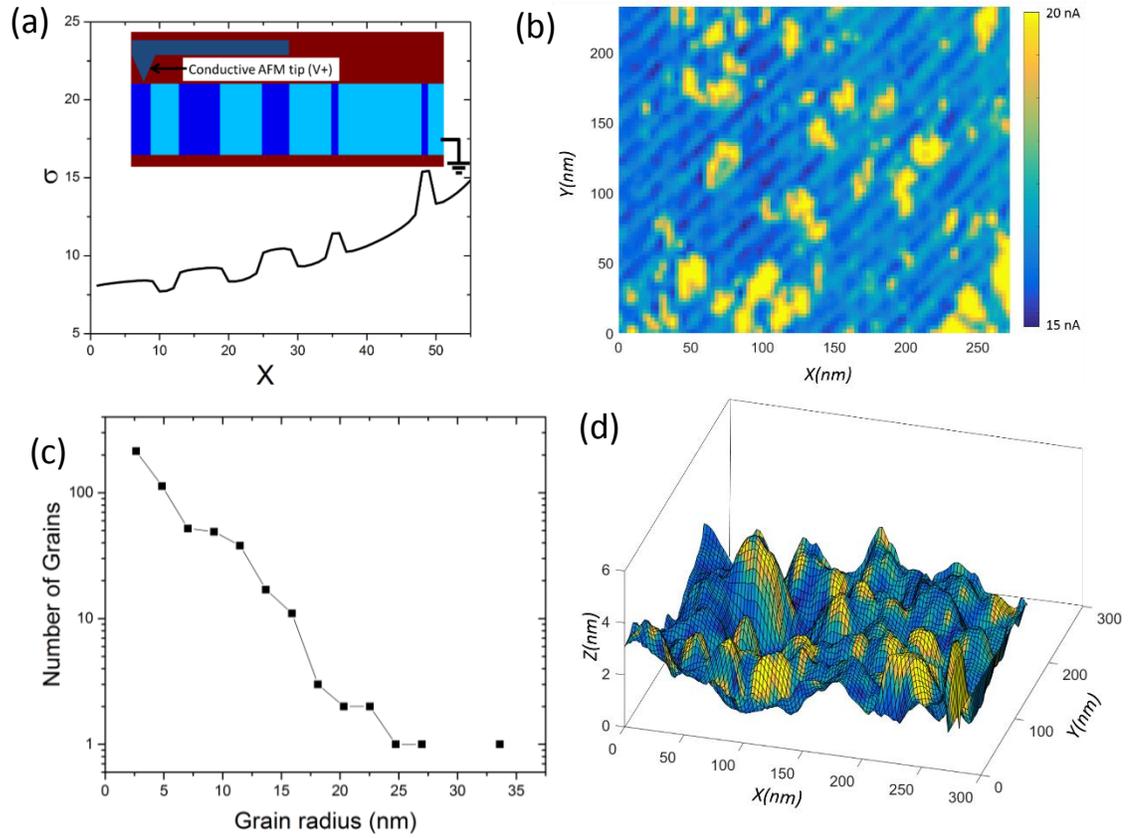

**Figure S3. Conductive AFM** (a) Inset shows a model granular film with columnar features similar to those observed in our system. Blue corresponds to low resistances, light blue corresponds to intermediate resistances, and red is insulating. The plot shows simulated conductance as the tip is dragged along these features. (b) A conductive AFM measurement of an Nb film with fixed bias voltage. Current peaks are visible with width comparable to that of the grains. (d) Histogram of conductive AFM grains extracted with an object finder. (c) A 3D plot with measured surface height shown in the Z axis and current information shown in color.

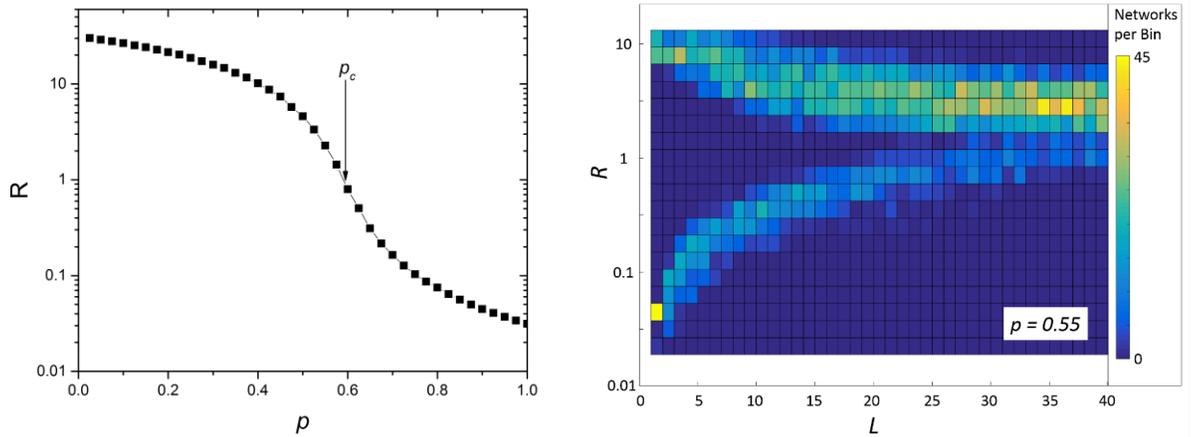

**Figure S4. Resistor Network Simulations** (a) The resistance of a random resistor model with a probability, p, of there being a low resistance connection rather that a high resistance connection. (b) A histogram of resistance as a function of network width, L. The islands split into two groupings with increasing L, one with lower resistance that's spanned by a low resistance network and another that's split by a hole in the network. The two groupings are equal in number near pc, but the higher resistance grouping is dominant for p>pc and the lower resistance grouping dominates p<pc.

### References(1-4)